\documentstyle[twoside,fleqn,times,epsfig]{actaps}

\begin{document}

\headings{1}{4}
\def\authorlist{K. Banaszek {\em et al.}}
\def\shorttitle{Determination of the Wigner function\ldots}

\title{DETERMINATION OF THE WIGNER FUNCTION FROM PHOTON STATISTICS}
\author{{K. Banaszek$^a$}\email{Konrad.Banaszek@fuw.edu.pl},
C. Radzewicz$^a$, K. W\'{o}dkiewicz$^{ab}$, J. S. Krasi\'{n}ski$^c$}
{(a) Wydzia{\l} Fizyki, Uniwersytet Warszawski, Ho\.{z}a~69,
PL--00--681~Warszawa, Poland\\
(b) Abteilung f\"ur Quantenphysik, Universit\"at Ulm, D-89081 Ulm, Germany\\
(c) Center for Laser and Photonics Research,
Oklahoma State University, Stillwater, OK~74078, USA}

\day{26 April 1999}

\abstract{We present an experimental realisation of the direct scheme for
measuring the Wigner function of a single quantized light mode. In this
method, the Wigner function is determined as the expectation value of the
photon number parity operator for the phase space displaced quantum state.}

\medskip


The Wigner function provides a complete representation of the quantum
state in the form that is analogous to a classical phase space probability
distribution \cite{WignPRA32}. An interesting and nontrivial problem is
how to relate the Wigner function to directly measurable quantities
so that it can be determined from data collected in a feasible
experimental scheme.  The first answer was given by Vogel and Risken
\cite{VogeRiskPRA89} who noted that marginal distributions of the Wigner
function can be measured with a balanced homodyne detector, and that the
inverse transformation is possible by tomographic back-projection. This
idea has been realized in a beautiful experiment by Smithey and co-workers
\cite{SmitBeckPRL93}, and it has quickly become a useful tool in studying
quantum statistical properties of optical radiation \cite{BreiSchiNAT97}.

In this contribution we briefly review our recent experimental realization
\cite{BanaRadzPRA99} of the direct scheme for measuring the Wigner
function of a light mode \cite{WallVogePRA96,BanaWodkPRL96}.  This method,
based on photon counting, provides the complete Wigner representation of
the quantum state without using any numerical reconstruction algorithms.

The basic idea underlying our experimental scheme is that the Wigner
function $W(\alpha)$ at a given phase space point $\alpha$ is itself a
well defined quantum observable. This observable can be represented as
the expectation value of the displaced photon number parity operator:
\begin{equation}
W(\alpha) = \frac{2}{\pi} \left\langle \hat{D}(\alpha) \right.
\sum_{n=0}^{\infty}(-1)^{n} |n\rangle\langle n| 
\left. \hat{D}^{\dagger}(\alpha) \right\rangle.
\end{equation}
Here $\hat{D}(\alpha) = \exp(\alpha \hat{a}^\dagger - \alpha^{\ast}
\hat{a})$ denotes the displacement operator, and 
$|n\rangle\langle n|$ are projections on Fock
states. The two elements of the above expression: the displacement
transformation and the projections on Fock states have a straightforward
optical realisation. The displacement transformation can be implemented
by interfering the signal field at a nearly fully transmitting beam
splitter with a probe coherent field, and projections on Fock states
are given by the photon statistics.

\begin{figure}
\centerline{\epsfig{file=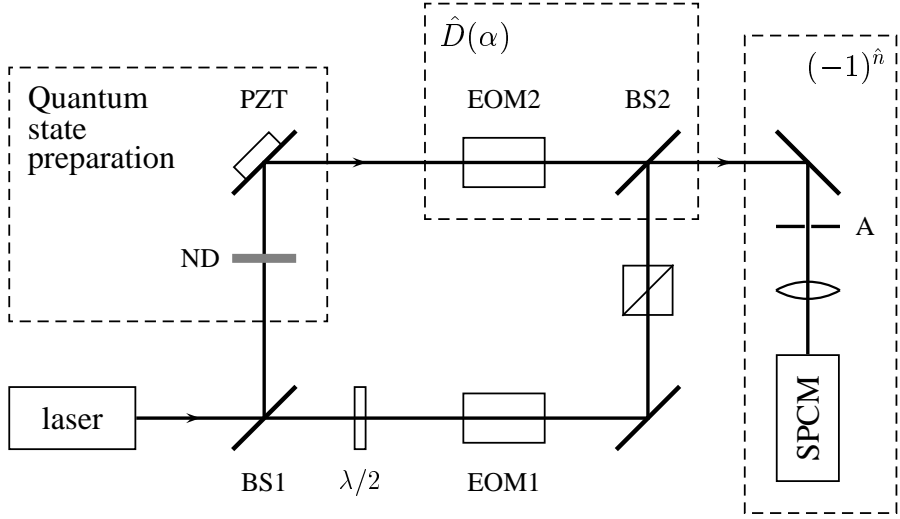,width=12cm,%
bbllx=110,bblly=415,bburx=510,bbury=650}}

\bigskip

\caption{Fig. 1. The experimental setup for measuring the Wigner
function of a single light mode. BS1 and BS2 are quartz plates serving
as high-transmission beam splitters. The quantum state to be measured
is prepared using the neutral density filter ND and the mirror mounted
on a piezoelectric translator PZT. The electrooptics modulators EOM1
and EOM2 define the amplitude and the phase of the displacement
$\hat{D}(\alpha)$. The quantum state, after applying the displacement
transformation, is detected with the single photon counting module
SPCM.}
\end{figure}

Our experimental scheme, presented in Fig.~1, is constructed as  an
unbalanced Mach-Zender interferometer with the beams in the two arms
serving as the signal and the probe fields. The source of light is an
attenuated beam from a frequency-stabilized single-mode He:Ne laser.
The quantum state is prepared as a weak coherent state using the neutral
density filter ND. Additionally, the mirror mounted on a piezoelectric
translator PZT can be used to generate a statistical mixture of coherent
states with fluctuating phase.

The displacement transformation $\hat{D}(\alpha)$ is realized using the
phase modulator EOM2 and the high-transmission beam splitter BS2 whose
second input port is fed with a coherent probe beam. The modulator EOM2
performs rotation in the phase space, while the probe beam effectively
shifts the phase space in a fixed direction. The value of the shift is
proportional to the probe beam amplitude, which is controlled in the
setup using the Pockels cell EOM1 placed between the half-wave plate
and a polarizer. Thus, the point of the phase space at which the Wigner
function is measured, is defined in our experiment by voltages applied
to the modulators EOM1 and EOM2.

The photon statistics of the displaced signal field is measured using
an avalanche photodiode operated in the single photon counting module.
The count rate is adjusted to the level such that a negligible fraction
of photons is missed due to the detector dead time. The experiment is
controlled by a computer, which collects photon statistics on a polar
grid in the phase space.

\begin{figure}
\begin{center}
\setlength{\unitlength}{.8in}
\begin{picture}(3,2.5)(0,3.5)
\put(3,5.8){\makebox(0,0)[rb]{\large (a)}}
\put(0,3.5){\epsfig{file=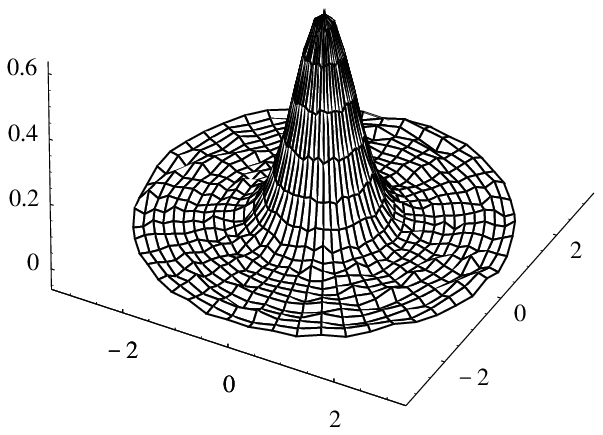,width=2.4in}}
\put(1.05,3.55){\makebox(0,0){Re($\beta$)}}
\put(2.9,3.9){\makebox(0,0){Im($\beta$)}}
\put(0,5.6){\makebox(0,0)[lb]{${\cal P}(\beta)$}}
\end{picture}
\hspace{.5cm}
\begin{picture}(3,2.5)(0,3.5)
\put(3,5.8){\makebox(0,0)[rb]{\large (b)}}
\put(0,3.5){\epsfig{file=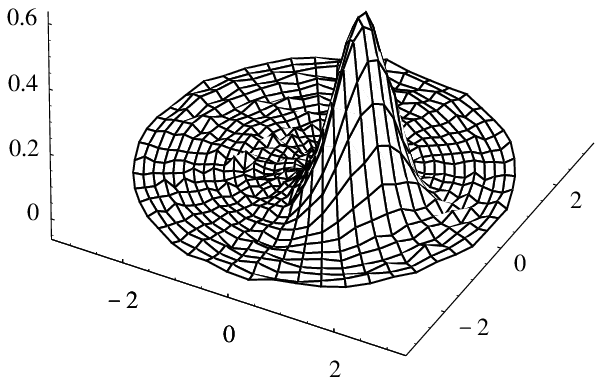,width=2.4in}}
\put(1.05,3.55){\makebox(0,0){Re($\beta$)}}
\put(2.9,3.9){\makebox(0,0){Im($\beta$)}}
\put(0,5.6){\makebox(0,0)[lb]{${\cal P}(\beta)$}}
\end{picture}

\vspace{.5cm}

\begin{picture}(3,2.5)(0,3.5)
\put(3,5.8){\makebox(0,0)[rb]{\large (c)}}
\put(0,3.5){\epsfig{file=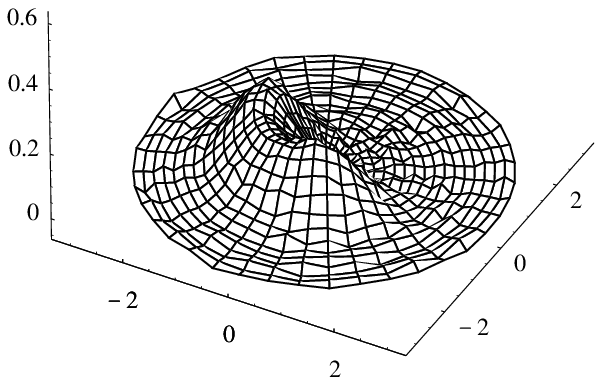,width=2.4in}}
\put(1.05,3.55){\makebox(0,0){Re($\beta$)}}
\put(2.9,3.9){\makebox(0,0){Im($\beta$)}}
\put(0,5.6){\makebox(0,0)[lb]{${\cal P}(\beta)$}}
\end{picture}
\end{center}

\bigskip

\caption{Fig. 2. The measured Wigner functions for (a) the vacuum, (b)
a weak coherent state with approximately one photon, and (c) a phase
diffused coherent state. The photon statistics was collected on a polar
grid spanned by 20 amplitudes, and 50 phases for the plots (a) and (b),
or 40 phases for the plot (c). The duration of a single counting
interval was 40~$\mu$s for (a) and (b) and 30~$\mu$s for (c). The
measurements were performed for slightly different
laser intensity, and the radial coordinate for each of the graphs
was scaled separately.}
\end{figure}

In Fig.~2 we present experimental results obtained for the vacuum state,
a coherent state, and a phase diffused coherent state.  The scan for
the vacuum state has been performed with the blocked signal path,
and the phase diffused coherent state has been generated by applying a
400~Hz sine waveform to the piezoelectric translator.  At each point of
the grid, the photon statistics has been collected from 8000 counting
intervals. The graphs are parameterized with the
complex variable $\beta=n_{\mbox{\scriptsize vac}}^{1/2}e^{i\phi}$, where
$n_{\mbox{\scriptsize vac}}$ is the average number of photons registered
for the blocked signal path, and $\phi$ is the phase shift generated by
the modulator EOM2. The photon statistics $p_n(\beta)$ collected at a
point $\beta$ is processed to yield
\begin{equation}
{\cal P}(\beta) = \frac{2}{\pi}\sum_{n} (-1)^{n} p_{n}(\beta).
\end{equation}
In the ideal, loss-free limit this quantity is equal to the Wigner
function of the signal field. In a realistic case, ${\cal P}(\beta)$
can be related to a generalized $s$-ordered quasidistribution function
$W(\alpha;s)$:
\begin{equation}
\label{Eq:realistic}
{\cal P}(\beta) = \frac{1}{\eta T} W \left( 
\frac{\beta}{\sqrt{\eta T}} ; - \frac{1-\eta T}{\eta T} \right),
\end{equation}
where $\eta$ is the quantum efficiency of the photodetector and $T$ is
the power transmission of the beam splitter performing the phase space
displacement. The right-hand side of Eq.~(\ref{Eq:realistic}) can be
interpreted as the Wigner function of the signal field that has passed
through a dissipation process with the losses characterized by the
overall efficiency $\eta T$. For our setup, the efficiency of the
photon counting module specified by the manufacturer is $\eta=70\%$,
and the power transmission of the beam splitter BS2 is $T=98.6\%$.
This gives the value of the ordering parameter equal to
$s=1-1/\eta T = -0.45$.

Further details and discussion of various aspects of the
experiment can be found in Ref.~\cite{BanaRadzPRA99}.
An important factor that should be taken into account in the analysis of
experimental data is the mode-mismatch between the fields interfering at
the beam splitter BS2. The effect of the mode-mismatch is quite different
from balanced homodyne detection, where it can be included in the overall
efficiency parameter. In the photon counting scheme, the mode-mismatch
generates a slowly decaying gaussian envelope centered at the phase
space origin, which multiplies the Wigner function of the signal field.

\medskip

\noindent{\bf Acknowledgements} 
This research is supported by the KBN grant
2P03B~002~14. K.W.\ thanks the Alexander von Humboldt
Foundation for generous support and Prof.~W.~P.~Schleich
for his hospitality in Ulm.

\end{document}